\documentclass[ reprint,aps]{revtex4-2}
\usepackage{amsfonts}
\usepackage{eurosym}
\usepackage{amssymb}
\usepackage{amsmath}
\usepackage{tabularx,epsfig,color,tabularx,epstopdf}
\usepackage{graphicx}
\usepackage{dcolumn}
\usepackage{bm}

\usepackage{epstopdf}

\begin{document}

\title{Nonreciprocity-Enabled Chiral Stability of Nonlinear Waves}
\author{Wen-Rong Sun$^{1*}$} 
%\email{sunwenrong@ustb.edu.cn}
\author{Jesús Cuevas Maraver$^{2}$}
%\author{Boris A. Malomed$^{2,3}$}
\affiliation{$^{1}$ School of Mathematics and Physics, University of Science
and Technology Beijing, Beijing 100083, China\\
$^{2}$ Grupo de Física No Lineal, Departamento de Física Aplicada I, Universidad de Sevilla, Escuela Politécnica Superior, C/ Virgen de África, 7, 41011 Sevilla, Spain and Instituto de Matemáticas de la Universidad de Sevilla (IMUS), Edificio Celestino Mutis, Avda. Reina Mercedes $s / n, 41012$ Sevilla, Spain
}
%$^{3}$Instituto de Alta Investigaci\'{o}n,
%Universidad de Tarapac\'{a}, Casilla 7D, Arica, Chile
%}

\begin{abstract}
The control of wave propagation, particularly the quest for unidirectional transport, plays an important role in photonics and metamaterial science. While nonreciprocity is known to enable unidirectional amplification and stabilize complex solitons, its fundamental impact on the intrinsic stability of the nonlinear waves remains an open frontier. By obtaining exact analytical solutions for a dissipative, nonreciprocal sine-Gordon model (that was proposed in [\emph{Nature} 627, 528-533, 2024] and captures all essential physics of active materials) and performing a comprehensive stability analysis, we discover a chiral stability criterion: waves propagating in one direction exhibit robust stability, while their mirror-image counterparts are intrinsically unstable. This direction-dependent stability is confirmed by full nonlinear simulations.  Our findings identify the symmetry-breaking role of nonreciprocity in defining the stability of nonlinear waves, a key step toward controlling wave propagation in nonreciprocal media.
\end{abstract}

\maketitle


%\preprint{APS/123-QED}

% Force line breaks with \\%




%\tableofcontents

\section{Introduction} 

Nonreciprocity occurs in diverse fields and systems, inluding active matter~\cite{nonre1,nonre2,nonre3,nonre4}, quantum many-body systems~\cite{nonre41,nonre42,nonre43}, non-equilibrium systems~\cite{nonre5,nonre6,jonas},  neural networks~\cite{nonre7,nonre8}, and metamaterials~\cite{nonre9,nonre10,nonre11,nonre12,nonre13}.
It generally originates from broken symmetry or non-Hermiticity in the linear dynamics~\cite{nonre14,nonre15}, leading to a non-symmetric transfer function and unidirectional amplification. This amplification often drives the system into a nonlinear regime where coherent structures such as solitons can emerge. Recent studies of nonlinear nonreciprocal dynamics have revealed a variety of phenomena, ranging from pattern formation~\cite{nonre16,nonre17,nonre19} and nontopological nonlinear excitations~\cite{jonas2} to topological solitons~\cite{jonas}, topological defects~\cite{nonre18}, and topological edge-localized structures~\cite{nonre20,nonre21}.

A key platform for studying these effects is nonreciprocal active matter, characterized by local, non-reciprocal, and non-conservative interactions. The nonreciprocal sine-Gordon equation [Eq.~(\ref{sgnd})], which captures the essential physics of such systems, was introduced in Refs.~\cite{jonas,jonas2} and has successfully described experimental observations. However, due to the presence of dissipative and nonreciprocal terms, solving this model directly is challenging. Previous studies~\cite{jonas,jonas2} thus relied on analytical solutions of the integrable reciprocal limit—where $\eta=0$ and $\Gamma=0$ in Eq.~(\ref{sgnd})—combined with perturbation theory to explore topological and non-topological nonlinear excitations.
Notably, Ref.~\cite{jonas2} showed that nonreciprocity can stabilize breathing solitons by unidirectionally driving the constituent kink and antikink of a breather in the same direction. This stabilization requires a delicate balance among nonreciprocal gain, dissipation, and initial amplitude, limiting it to a narrow parameter region near an unstable fixed point. In a related study, Ref.~\cite{jonas} introduced a local non-reciprocal coupling mechanism that unidirectionally drives both solitons and antisolitons.

These results prompt a more fundamental question: how does nonreciprocity intrinsically influence the propagation and stability of nonlinear waves, beyond its role as an external driver or stabilizer of pre-existing nonlinear structures?  To address this question, we move beyond perturbative treatments and instead derive the fundamental modes (i.e., the exact analytical solutions)  of the nonreciprocal system described by Eq.~(\ref{sgnd}). These exact analytical solutions serve as elementary building blocks for understanding the system's dynamics through a systematic stability analysis.

In this paper, we report novel exact analytical solutions, including traveling periodic waves and kinks, for the dissipative nonreciprocal sine-Gordon model, which is notably non-integrable. Subsequently, a linear stability analysis of these traveling wave solutions leads to our key finding: the emergence of a chiral stability criterion, whereby the stability of a wave is fundamentally determined by its direction of propagation relative to the nonreciprocity parameter $\eta$. Furthermore, these results are further corroborated by direct numerical simulations of the full nonlinear dynamics, which confirm the chiral stability predicted by the linear stability analysis. Overall, our findings show that nonreciprocity does not merely drive or sustain complex structures, but breaks the symmetry of the dynamical landscape, thereby governing which elementary waves can stably exist.

\section{The continuum model, exact solutions and spectral problem}  

We study a nonreciprocal and dissipative sine-Gordon equation~\cite{jonas,jonas2}
\begin{equation}
\varphi_{t t}-\varphi_{x x}+\sin (\varphi)=-\eta \varphi_x-\Gamma \varphi_t, \label{sgnd}
\end{equation}
which represents a continuum limit of the nonreciprocal Frenkel–Kontorova model and captures the essential physics of active nonreciprocal materials~\cite{jonas,jonas2}. Here, $\varphi(x,t)$ is a real function of $x$ and $t$, while $\Gamma$ and $\eta$ denote the on-site damping and the nonreciprocity strength, respectively. The presence of the $\eta$  breaks spatial inversion symmetry.

The left-hand side of Eq.~(\ref{sgnd}) corresponds to the integrable sine-Gordon model, where two counter-propagating periodic waves of equal speed are known to share the same stability behavior~\cite{bernard, miller}. In this work, we systematically investigate how nonreciprocity alters such stability properties. The first challenge is to derive analytical solutions for this non-integrable system, which we address in the following.

To obtain the traveling wave solutions of Eq.~(\ref{sgnd}), by introducding $z=x-ct$, $\tau=t$ and $\phi(z,\tau)=\varphi(x,t)$, we have 
\begin{equation}
\left(c^2-1\right) \phi_{z z}-2 c \phi_{z \tau}+\phi_{\tau \tau}+\sin (\phi)=(c\Gamma-\eta)\phi_{z}-\Gamma\phi_{\tau}. \label{sgnd2}
\end{equation}
For subsequent discussion we assume
\begin{equation}
c=\eta/\Gamma\neq1,
\end{equation}
indicating that the propagation direction is determined by the ratio of nonreciprocity to damping. 

We aim to look for the stationary solutions of Eq.~(\ref{sgnd2}) by letting $\phi(z,\tau)=\psi(z)$. Integrating once leads to the following equation:
\begin{equation}
\frac{1}{2}\left(c^2-1\right) \psi^{\prime}(z)^2+1-\cos (\psi(z))=E,
\end{equation}
where $c=\eta/\Gamma\neq1$ and $E$ denotes the total energy.
This leads to four families of periodic traveling wave solutions of the form
\begin{equation}
\cos (\psi(z))=A+B \operatorname{sn}^2(\lambda z, k),\label{gs1}
\end{equation}
or 
\begin{equation}
\cos (\varphi(x,t))=A+B \operatorname{sn}^2(\lambda (x-ct), k),\label{gs}
\end{equation}
with $\operatorname{sn}$ being the Jacobi elliptic function of modulus $k$, and the parameters specified as follows:\\
$\bullet$ subluminal, rotational: $\quad A=-1, \quad B=2,
\lambda=\sqrt{\frac{2-E}{2\left(1-c^2\right)}}, \quad k=\sqrt{\frac{2}{2-E}}, \quad 0 \leq|\frac{\eta}{\Gamma}|<1, \quad E<0$.\\
$\bullet$ superluminal, rotational: $\quad A=1, \quad B=-2, \quad \lambda=\sqrt{\frac{E}{2\left(c^2-1\right)}}, \quad k=\sqrt{\frac{2}{E}},\quad |\frac{\eta}{\Gamma}|>1, \quad E>2$.\\
$\bullet$ subluminal, librational: $\quad A=-1, \quad B=2-E, \quad \lambda=\sqrt{\frac{1}{1-c^2}}, \quad k=\sqrt{\frac{2-E}{2}}, \quad 0 \leq|\frac{\eta}{\Gamma}|<1, \quad 0<E \leq 2$.\\
$\bullet$ superluminal, librational: $\quad A=1, \quad B=-E, \quad \lambda=\sqrt{\frac{1}{c^2-1}}, \quad k=\sqrt{\frac{E}{2}},\quad |\frac{\eta}{\Gamma}|>1, \quad  0<E \leq 2$.

The period of these solutions is $l = 2K(k)/\lambda$, where $K(k)$ is the complete elliptic integral of the first kind. In the limit $k \to 1$, the periodic waves degenerate into kink solutions. For subluminal waves ($E \to 0$),
\begin{equation}
\cos (\varphi(x,t))=-1+2 \tanh ^2\left(\frac{x-ct}{\sqrt{1-c^2}}\right),\label{k1}
\end{equation}
and for superluminal waves ($E \to 2$), 
\begin{equation}
\cos (\varphi(x,t))=1-2 \tanh ^2\left(\frac{x-ct}{\sqrt{1-c^2}}\right).\label{k2}
\end{equation}

To analyze the linear stability of the solutions given by Eq.~\eqref{gs1}, we substitude a perturbed ansatz into Eq.~\eqref{sgnd2}:
\begin{equation}
\phi(z, \tau)=\psi(z)+\epsilon w(z, \tau)+\mathcal{O}\left(\epsilon^2\right),
\end{equation}
where $\epsilon \ll 1$. By defining $w_1 = w$ and $w_2 = w_\tau$, the linearized dynamics can be expressed as 
\begin{equation}
\frac{\partial}{\partial \tau}\binom{w_1}{w_2}=\mathcal{L}\binom{w_1}{w_2},
\end{equation}
with the linear operator
\begin{equation}
\mathcal{L}=\left(\begin{array}{cc}
0 & 1 \\
-\left(c^2-1\right) \partial_z^2-\cos (\psi(z)) & 2 c \partial_z-\Gamma
\end{array}\right).
\end{equation}
Seeking separable solutions of the form
\begin{equation}
\binom{w_1(z, \tau)}{w_2(z, \tau)}=e^{\lambda \tau}\binom{W_1(z)}{W_2(z)},
\end{equation}
we obtain the spectral problem
\begin{equation}
\lambda\binom{W_1}{W_2}=\mathcal{L}\binom{W_1}{W_2},\label{st}
\end{equation}
where $\lambda$ is the eigenvalue and $(W_1, W_2)^\intercal$ is the corresponding eigenfunction. A solution is spectrally stable if $\Re(\lambda) \leq 0$ for all eigenvalues; otherwise, an exponential instability occurs with growth rate $\Re(\lambda) > 0$.

\begin{figure}[tbp]
\centering
\includegraphics[height=115pt,width=121pt]{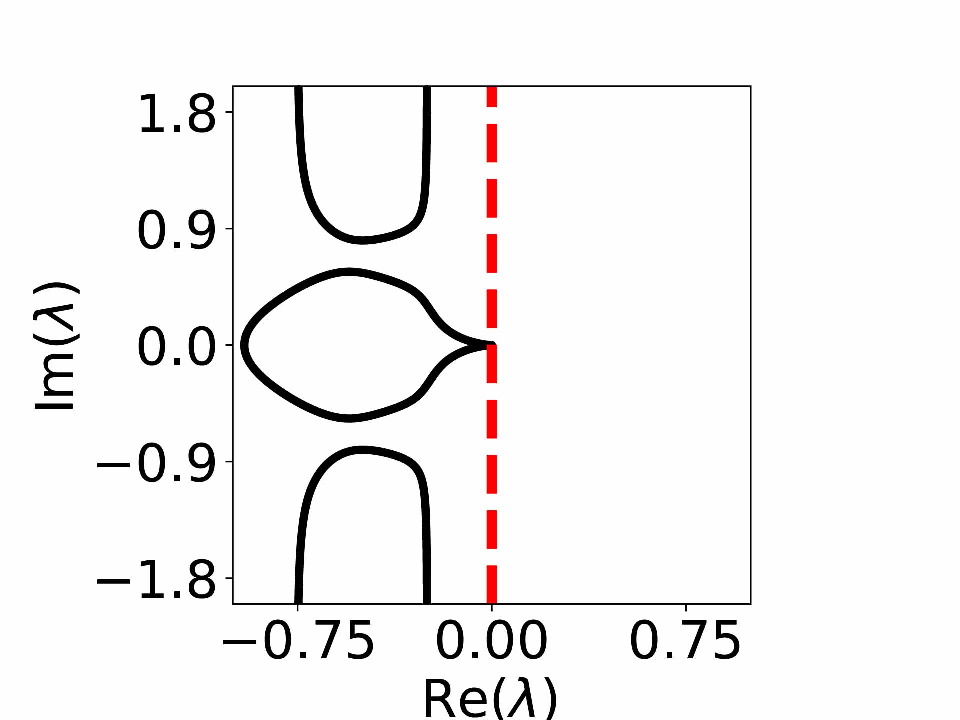} \includegraphics[height=115pt,width=121pt]{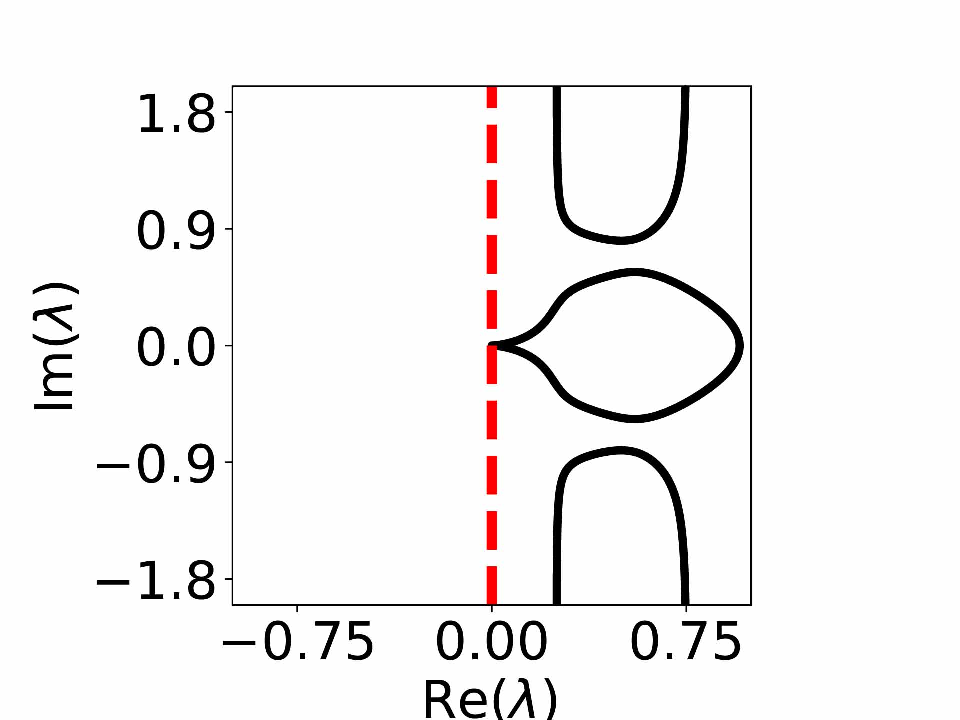}
\newline
\caption{The stability spectrum for the subluminal, rotational wave. The parameters are: $A=-1$, $B=2$, $\eta=0.5$, $E=-0.75$, $\Gamma=1$ (left), and $\Gamma=-1$ (right). }
\label{figg1}
\end{figure}

\begin{figure}[tbp]
\centering
\includegraphics[height=115pt,width=121pt]{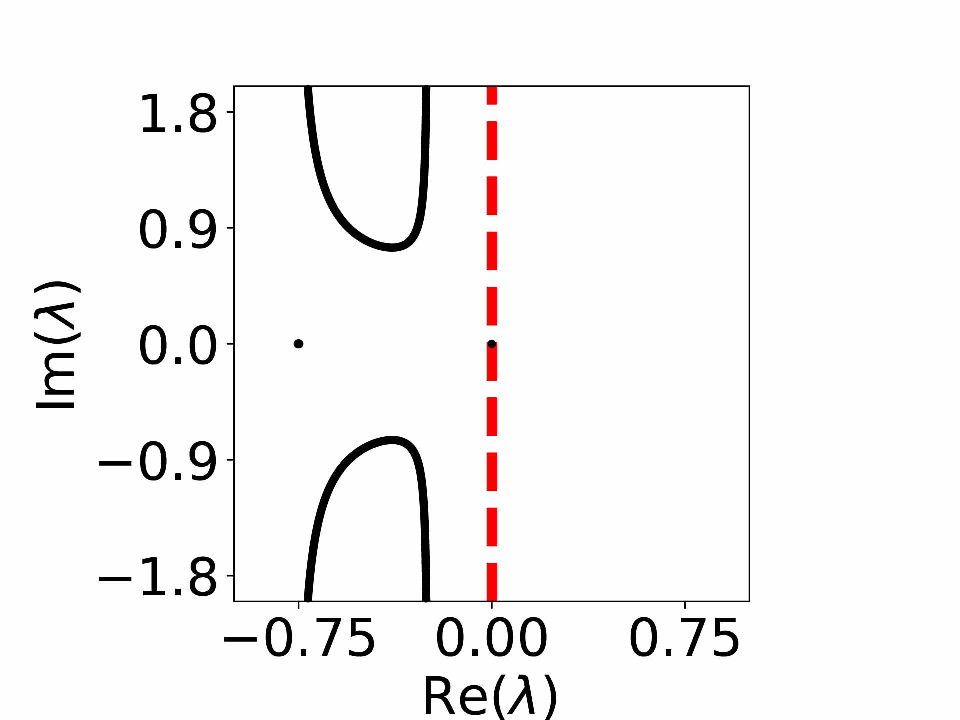} \includegraphics[height=115pt,width=121pt]{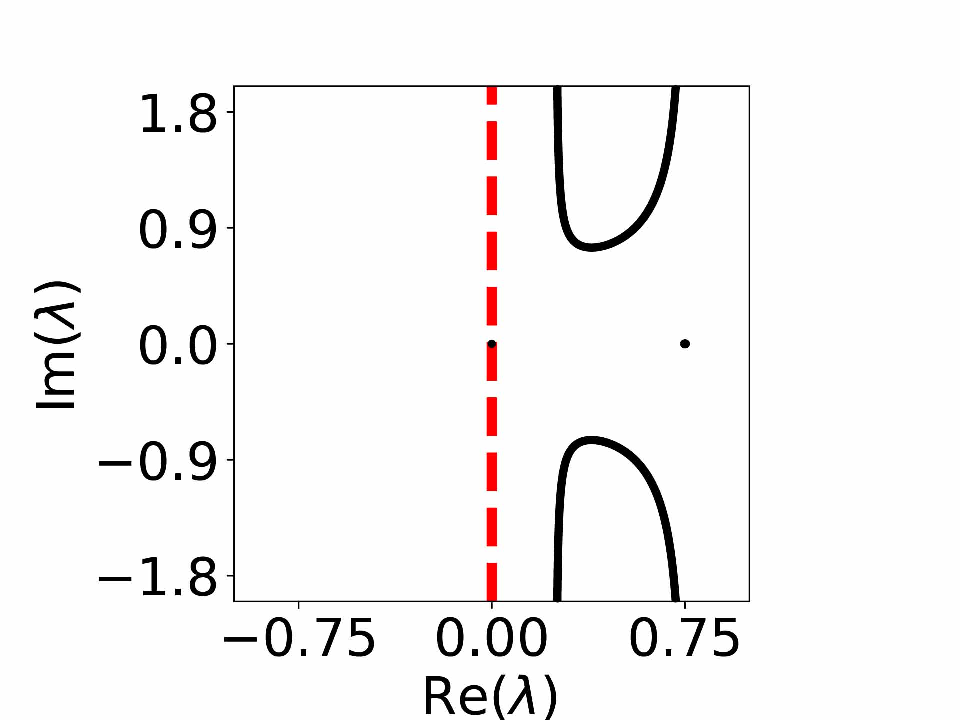}
\newline
\caption{The stability spectrum for the subluminal kink solution~(\ref{k1}). The parameters are:  $\eta=0.5$, $\Gamma=1$ (left), and $\Gamma=-1$ (right). }
\label{figg2}
\end{figure}

Since the coefficients of the spectral problem~\eqref{st} are periodic in $z$, we analyze it numerically using the Fourier-Floquet-Hill method~\cite{bernard2}. This method has been verified to be applicable to periodic operator problems in various systems~\cite{wr1}. By Floquet's theorem, the eigenmode components $W_1$ and $W_2$ admit the Bloch-wave expansion
\begin{eqnarray}
W_{1}(z) &=&e^{i\mu z}\sum_{n=-\infty }^{\infty }W_{1,n}\exp \left( \frac{%
2\pi }{pl}inz\right) ,  \notag \\
&&  \label{f1} \\
W_{2}(y) &=&e^{i\mu z}\sum_{n=-\infty }^{\infty }W_{2,n}\exp \left( \frac{%
2\pi }{pl}inz\right) ,  \notag
\end{eqnarray}%
where $p$ is a subharmonicity index, allowing the perturbation period $pl$ to be an integer multiple of the base solution period $l$, and $\mu \in [-\pi/(pl), \pi/(pl)]$ is the Floquet exponent (or quasi-momentum). $W_{1,n}$ and $W_{2,n}$ are the corresponding Fourier coefficients. This formulation enables a systematic assessment of the subharmonic (in)stabilities of the solutions~\eqref{gs}.

\section{Chiral stability of nonlinear waves}  
We illustrate these findings using subluminal rotational waves as an example. Fixing $\eta = 0.5$ and considering $\Gamma = \Gamma_{\pm 1} = \pm 1$, we examine two periodic waves propagating in opposite directions with speeds $c_1 = \eta/\Gamma_{+1}$ and $c_2 = \eta/\Gamma_{-1} = -c_1$. As shown in Fig.~\ref{figg1}, the wave with $c = c_1$ is spectrally stable against subharmonic perturbations, whereas its mirror image propagating in the opposite direction ($c = c_2$) exhibits modulational instability. This demonstrates that wave stability itself is chiral—that is, intrinsically tied to the direction of propagation.
This direction-dependent stability is a generic feature, equally evident in our analysis of kink solitons~(\ref{k1}). As shown in Fig.~\ref{figg2}, a kink traveling with speed 
$c=\frac{\eta}{\Gamma}$ is linearly stable, while its counter-propagating counterpart is unstable. The chiral stability mechanism thus applies universally—from periodic waves to localized kink solitons—governing different coherent structures in the nonreciprocal regime.

Chirality here represents a fundamental breaking of mirror symmetry: a wave traveling in one direction can form a robust, stable structure, while its mirror-symmetric counterpart is inherently unstable. This chiral stability reflects a dynamical selection mechanism governed by the nonreciprocity parameter $\eta$, whose sign acts as a master switch, unilaterally stabilizing waves in one direction while suppressing those in the other.

\begin{figure}[tbp]
\centering
\includegraphics[height=115pt,width=121pt]{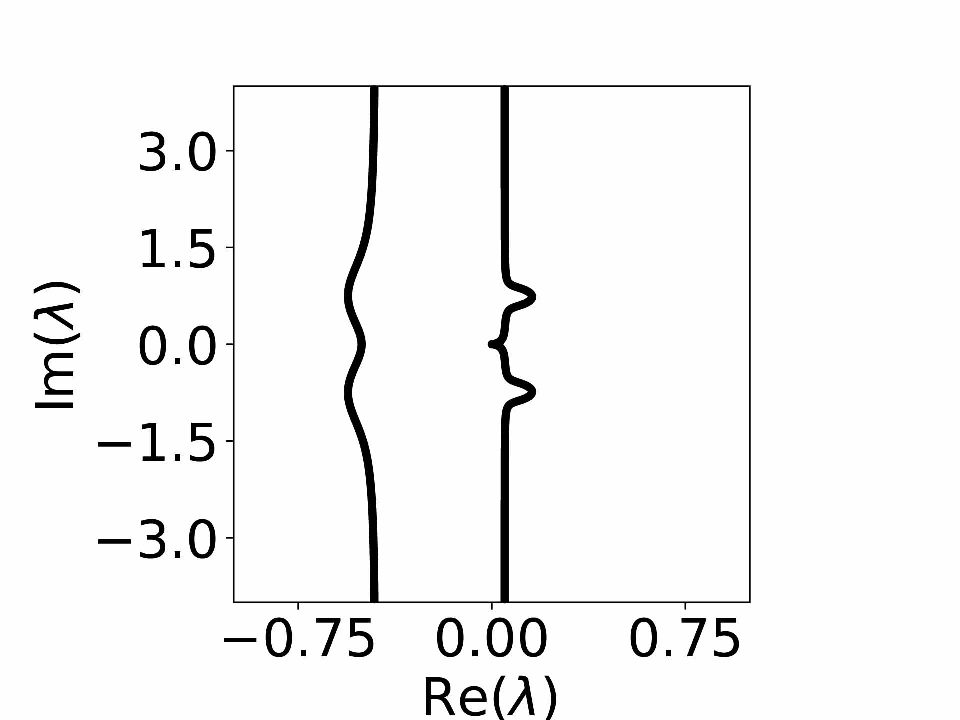} \includegraphics[height=115pt,width=121pt]{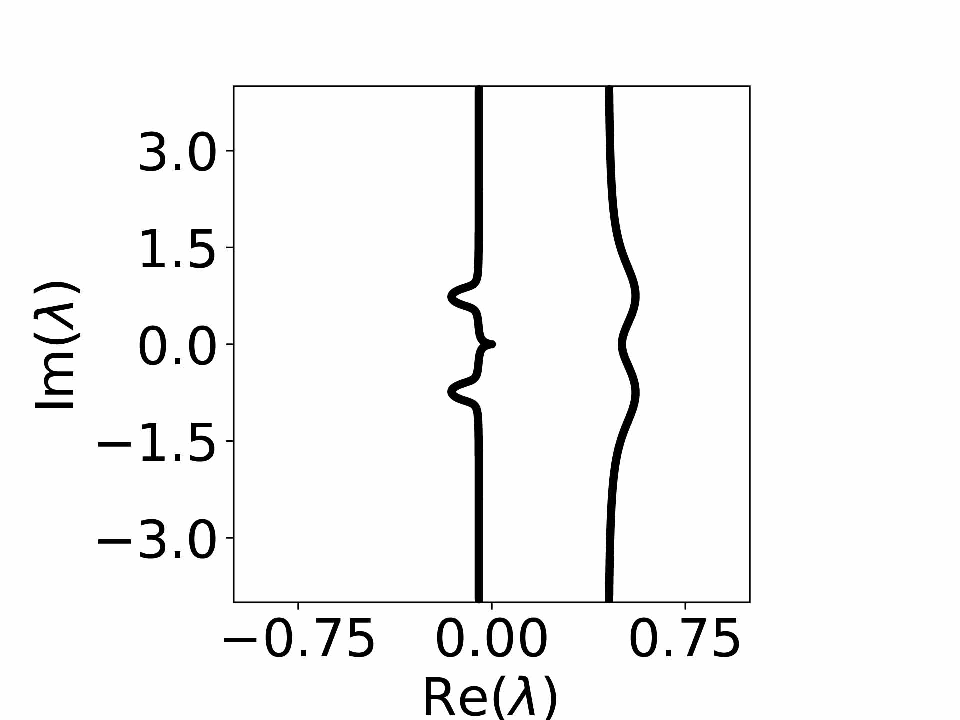}
\newline
\caption{The stability spectrum for the superluminal, rotational wave. The parameters are: $A=1$, $B=-2$, $\eta=0.5$, $E=3.1$, $\Gamma=0.4$ (left), and $\Gamma=-0.4$ (right). }
\label{figg3}
\end{figure}

\begin{figure}[tbp]
\centering
\includegraphics[height=115pt,width=121pt]{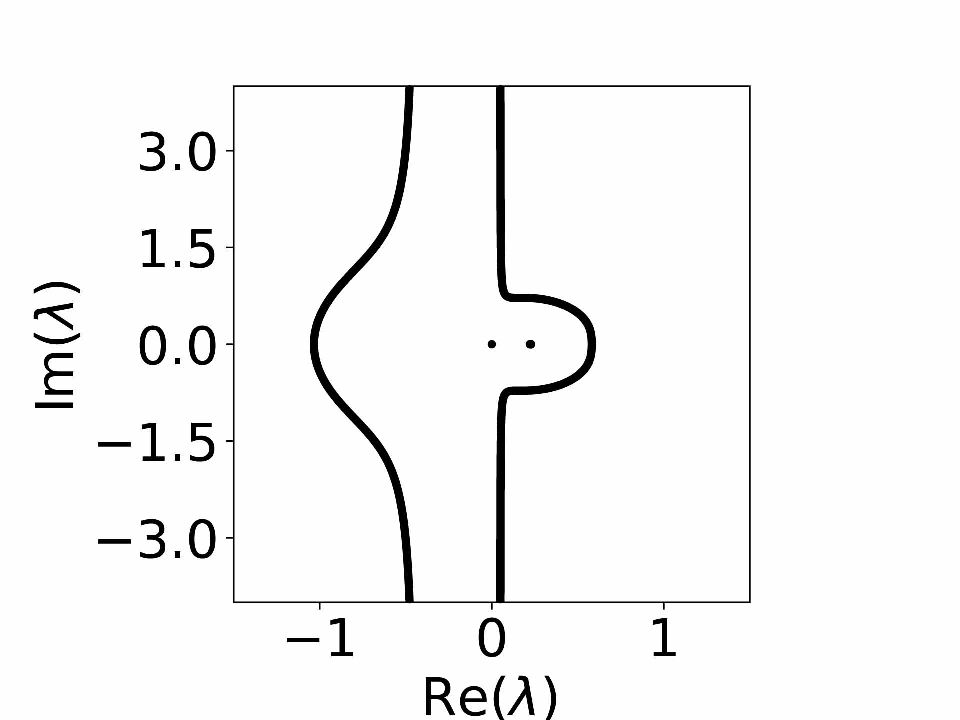} \includegraphics[height=115pt,width=121pt]{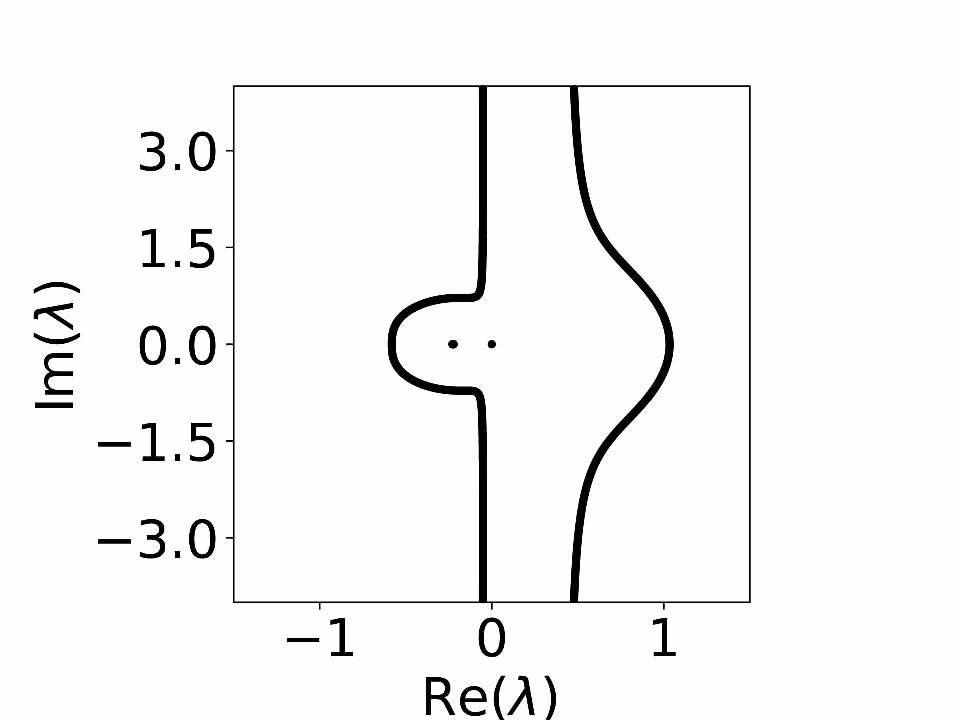}
\newline
\caption{The stability spectrum for the superluminal kink solution~(\ref{k2}). The parameters are: $\eta=0.5$, $\Gamma=0.4$ (left), and $\Gamma=-0.4$ (right). }
\label{figg4}
\end{figure}

For superluminal case, we take superluminal, rotational wave as the example. The effect of chiral symmetry breaking is not limited to a binary selection of stable modes. Even in parameter regimes where both counter-propagating waves are unstable, their growth rates differ significantly, as detailed in Figs.~\ref{figg3} and~\ref{figg4}. This shows that nonreciprocity affects the temporal dynamics of instability itself, favoring perturbations of one chirality over the other during transient evolution. This finding stands in stark contrast to the reciprocal case, where counter-propagating waves exhibit identical growth spectra. Nonreciprocity shapes the linear spectrum beyond a binary stability filter by structuring a hierarchy of growth rates.

\section{Nonlinear dynamics} We simulate the nonlinear dynamics of Eq.~\eqref{sgnd} using subluminal and superluminal rotational waves as initial conditions. These solutions reduce to kink-like structures in the limit $k \to 1$.

As $\cos^{-1}(x)$ is only defined at $[0,\pi)$, solution in (\ref{gs}) is not differentiable when the cosine is inverted. Because of this, for the particular choice of $A$ and $B$ made above, the subluminal and superluminal rotational waves can be written as

\begin{equation}
\varphi(x,t)=2\operatorname{am}(\lambda (x-ct), k)+\pi
\end{equation}
and
\begin{equation}
\varphi(x,t)=2\operatorname{am}(\lambda (x-ct), k), 
\end{equation}
respectively, with $\operatorname{am}$ being the Jacobi amplitude function of modulus $k$. In order to perform simulations, we took the initial conditions $\varphi(x,0)$ and $\partial_t\varphi(x,0)$ (with $\partial_t\varphi(x,t)=-2\lambda c\operatorname{dn}(\lambda (x-ct), k)$) with the parameters indicated in the footnote of Figs.~\ref{figg1}--\ref{figg4}. The choice of a domain $[-L,L]$ is important as we need that the topological charge $Q=\varphi(L,t)-\varphi(-L,t)$ is an integer multiple of $\pi$ in order to take adequate anti-periodic boundary conditions. Because of this, $2L=N\Lambda$, with $N\in\mathbb{N}$ and $\Lambda=2\mathbf{K}(k)/\lambda$ being the wavelength of the rotational wave, which implies $Q=2N\pi$ and the boundary conditions are taken as $\varphi(L+\delta x)=\varphi(L)+Q$ and $\varphi(-L-\delta x)=\varphi(L)-Q$, with $\delta x$ being the discretization of the finite differences (we have taken $\delta x=0.02\Lambda$. In the simulations that follow, we have taken $N=10$).

\begin{figure}[tbp]
\centering
\begin{tabular}{cc}
\includegraphics[width=.25\textwidth]{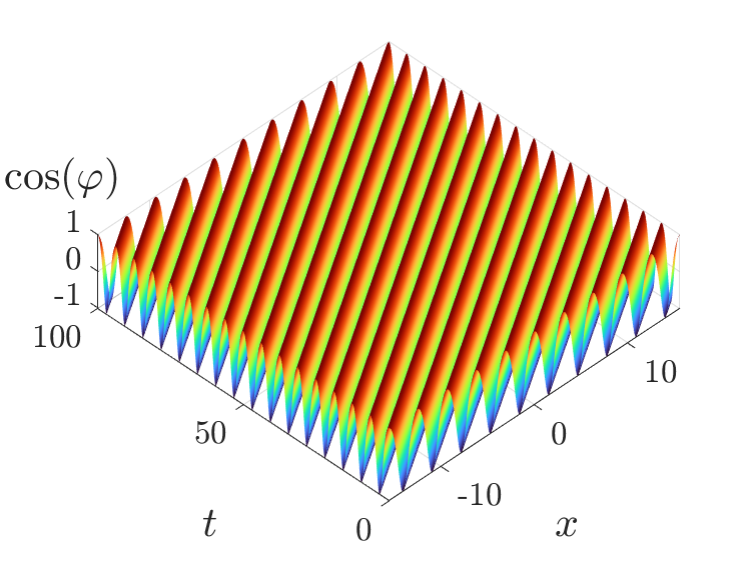} & 
\includegraphics[width=.25\textwidth]{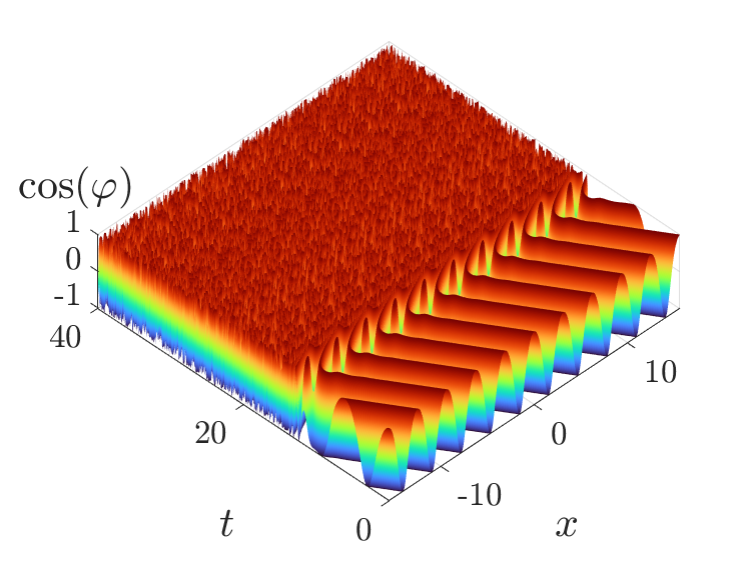} \\
\includegraphics[width=.25\textwidth]{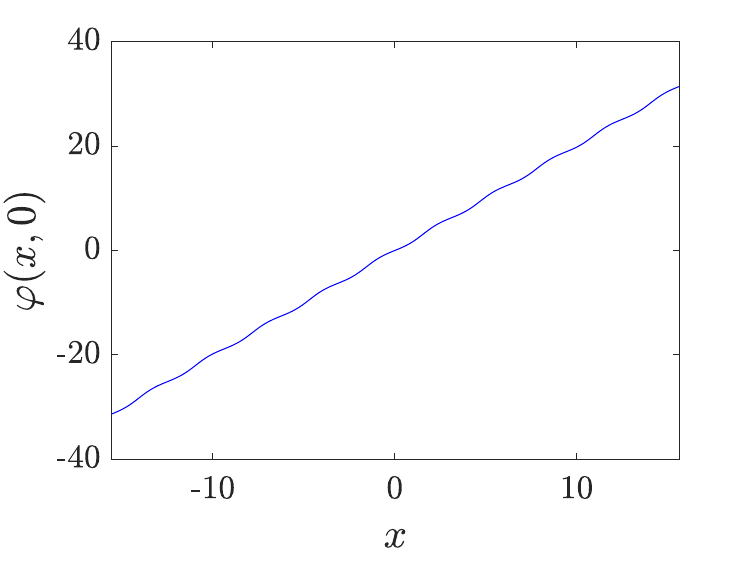} & 
\includegraphics[width=.25\textwidth]{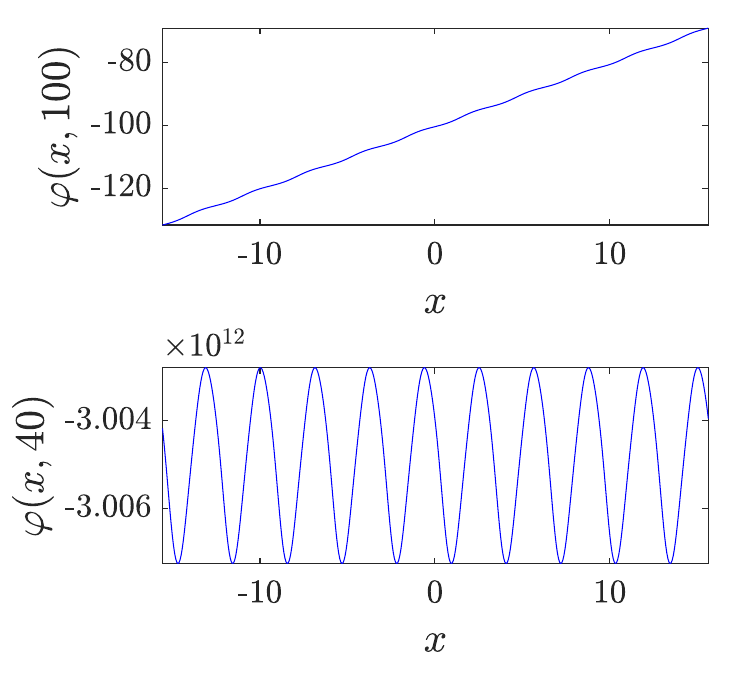} \\
\end{tabular}
\caption{(Top panels) Spatiotemporal evolution of $\cos(\varphi(x,t))$ for the subluminal rotational waves analysed in Fig.~\ref{figg1}. Left (right) panel corresponds to the solution with $c>0$ ($c<0$). Bottom left panel shows the initial condition $\varphi(x,0)$ for the simulation whereas right panel shows the profile at the end of the simulation for the $c>0$ (top) and the $c<0$ (bottom) case.}
\label{figg5}
\end{figure}

\begin{figure}[tbp]
\centering
\begin{tabular}{cc}
\includegraphics[width=.25\textwidth]{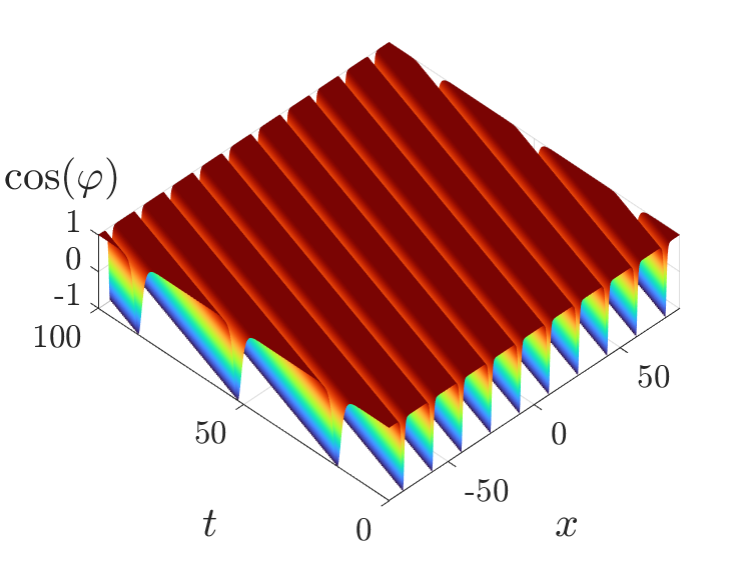} & 
\includegraphics[width=.25\textwidth]{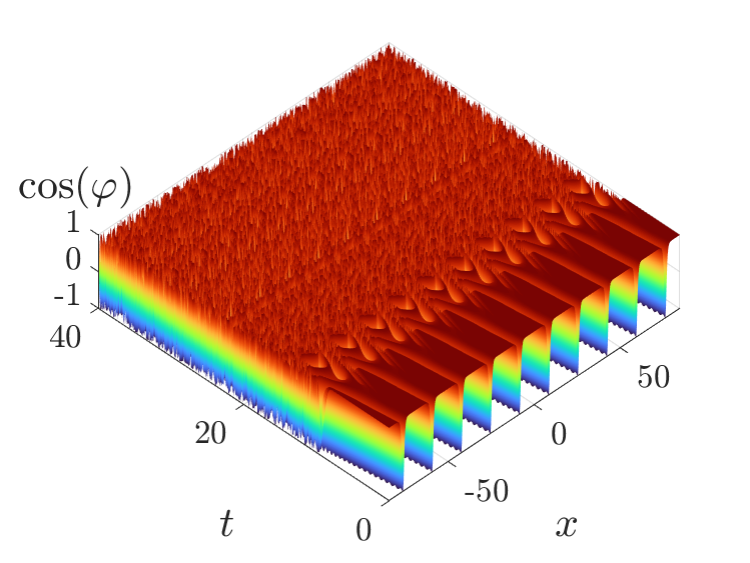} \\
\includegraphics[width=.25\textwidth]{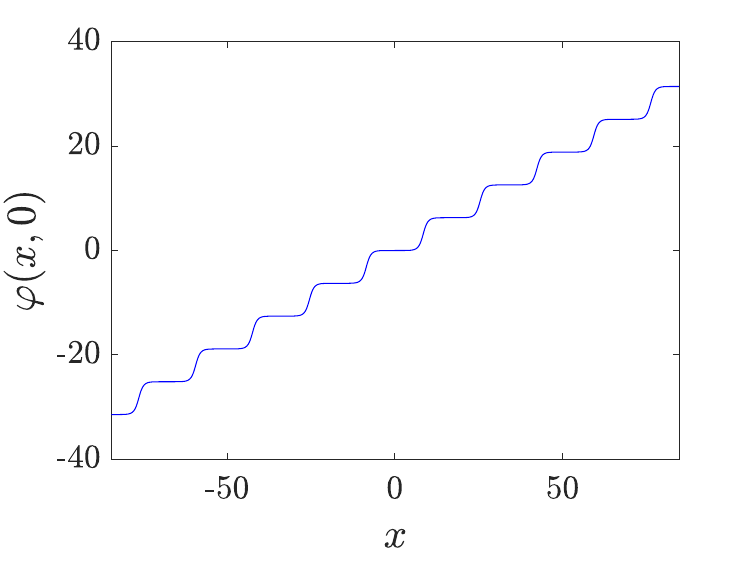} & 
\includegraphics[width=.25\textwidth]{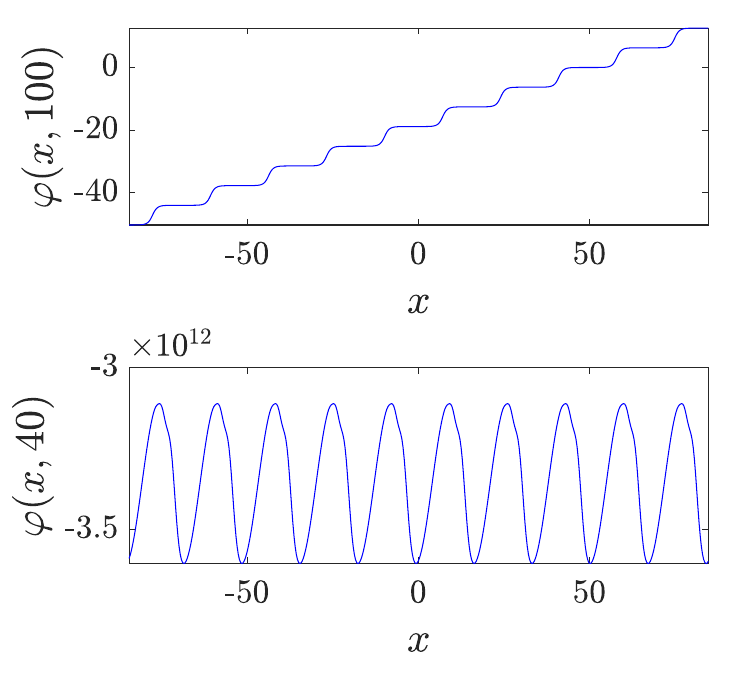} \\
\end{tabular}
\caption{Same as Fig.~\ref{figg5} but for the subluminal kink-like solutions~(\ref{k1}) corresponding to Fig.~\ref{figg2}.}
\label{figg6}
\end{figure}

\begin{figure}[tbp]
\centering
\begin{tabular}{cc}
\includegraphics[width=.25\textwidth]{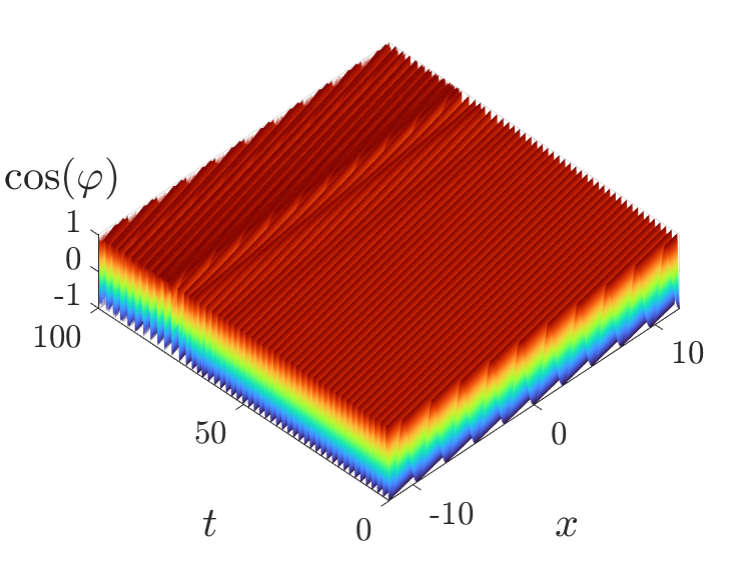} & 
\includegraphics[width=.25\textwidth]{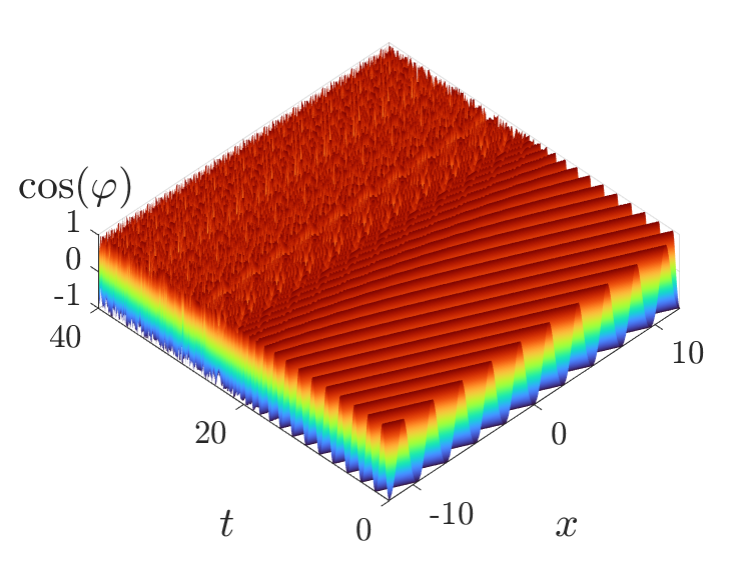} \\
\includegraphics[width=.25\textwidth]{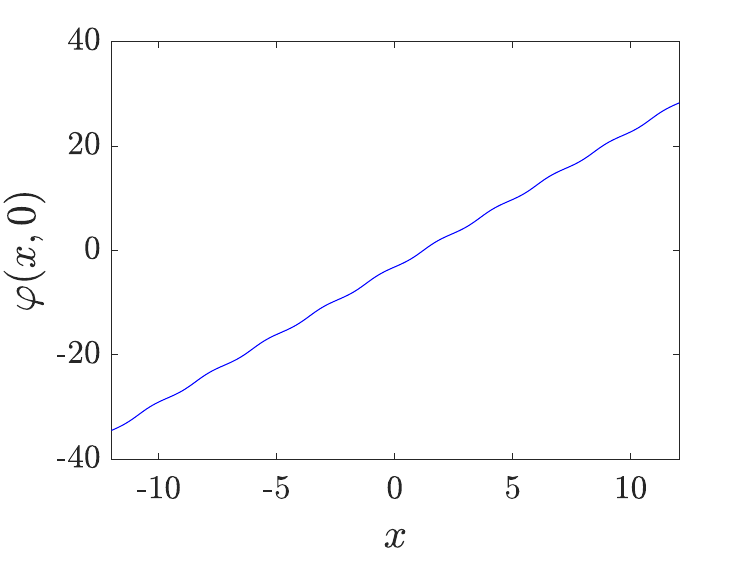} & 
\includegraphics[width=.25\textwidth]{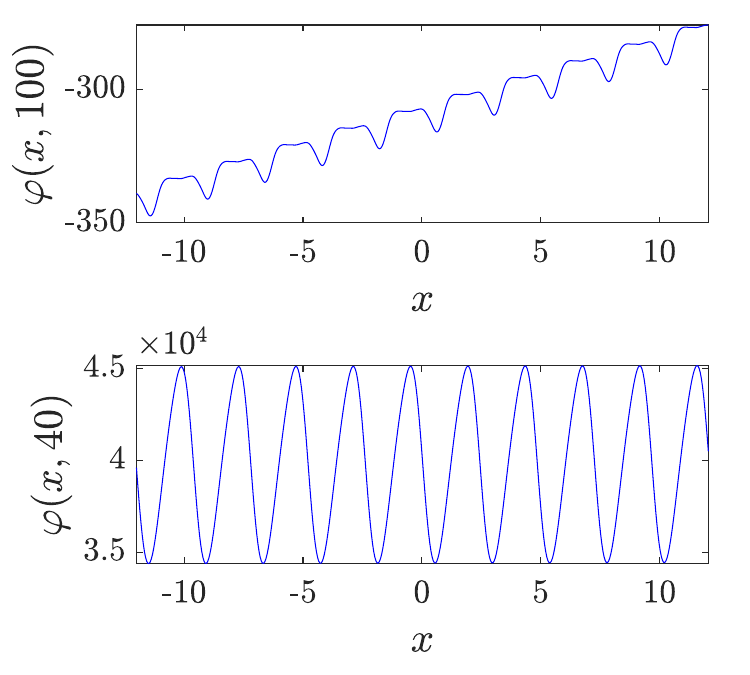} \\
\end{tabular}
\caption{(Top panels) Spatiotemporal evolution of $\cos(\varphi(x,t))$ for the superluminal rotational waves analysed in Fig.~\ref{figg3}. Left (right) panel corresponds to the solution with $c>0$ ($c<0$). Bottom left panel shows the initial condition $\varphi(x,0)$ for the simulation whereas right panel shows the profile at the end of the simulation for the $c>0$ (top) and the $c<0$ (bottom) case.}
\label{figg7}
\end{figure}

\begin{figure}[tbp]
\centering
\begin{tabular}{cc}
\includegraphics[width=.25\textwidth]{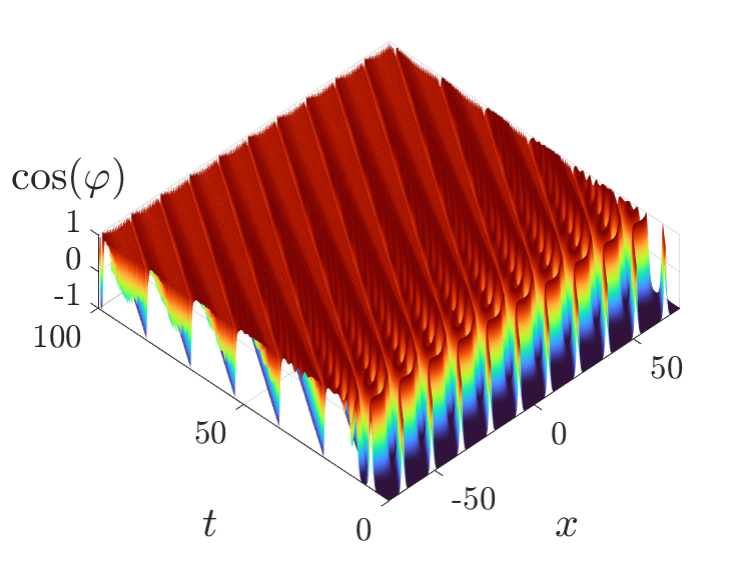} & 
\includegraphics[width=.25\textwidth]{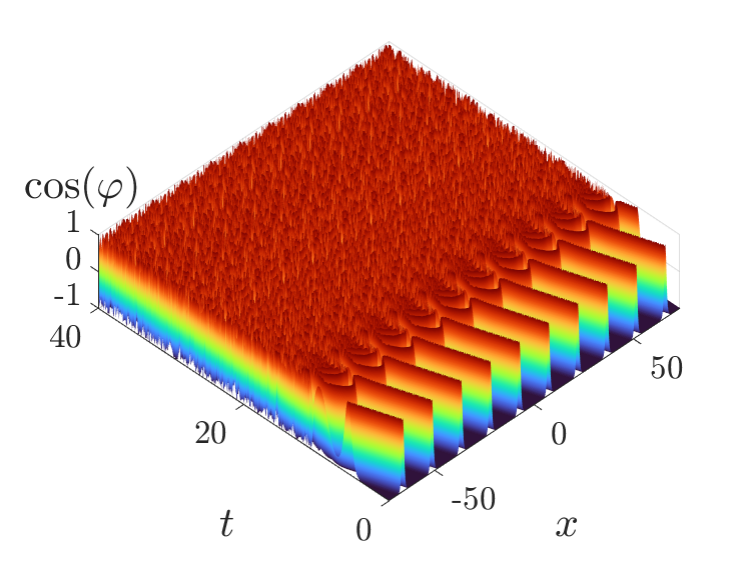} \\
\includegraphics[width=.25\textwidth]{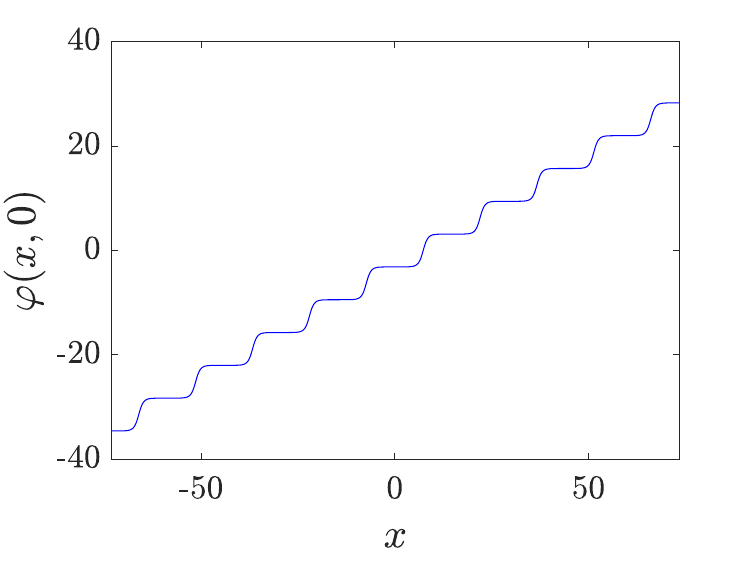} & 
\includegraphics[width=.25\textwidth]{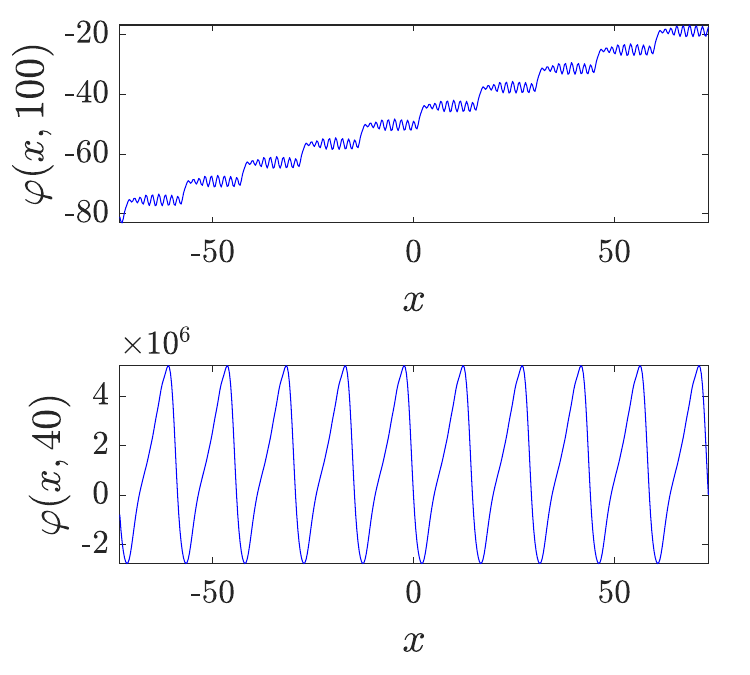} \\
\end{tabular}
\caption{Same as Fig.~\ref{figg7} but for the superluminal kink-like solutions~(\ref{k2}) corresponding to Fig.~\ref{figg4}.}
\label{figg8}
\end{figure}

As Fig.~\ref{figg5} shows, the subluminal rotational waves with $c>0$ are stable, whereas they are unstable for $c<0$ leading to the removal of the topological charge and to a huge increment of the amplitude. This scenario is similar to that of the kink shown in Fig.~\ref{figg6}. For superluminal waves (see Figs.~\ref{figg7} and \ref{figg8}), we observe a similar outcome to the subluminal ones when $c<0$; however, when $c>0$ the solutions are unstable but are topologically protected (that is, the topological charge does not change). Excellent agreement is observed between the nonlinear evolution and the predictions of the linear stability analysis.

\section{Conclusions.} This work has established a fundamental principle governing wave dynamics in nonreciprocal media: the chiral stability criterion. Through exact analytical solutions and spectral analysis of the dissipative nonreciprocal sine-Gordon model, we have demonstrated that wave stability becomes an intrinsic property of propagation direction. The emergence of this direction-dependent stability—where waves propagating in one direction exhibit robust stability while their mirror images are unstable—represents a breaking of dynamical symmetry.
Our findings indicate that nonreciprocity governs the stability of elementary wave components, in addition to driving and stabilizing specific nonlinear structures. This behavior is observed across both periodic waves and topological kink solitons, suggesting it may be a general feature of nonreciprocal systems. 
In our future work, we plan to investigate the role of nonreciprocity in modulational instability (MI). While MI is a fundamental mechanism generating localized structures like solitons and rogue waves in reciprocal media, it remains unknown how nonreciprocal driving alters its dynamics. A central goal is to determine whether such nonreciprocal MI can lead to localized structures with broken symmetry.

%\begin{acknowledgments}
\textit{Acknowledgments.} The work of Sun is supported by the NSFC (Grants No. 12575001) and the Open Project of Key Laboratory of Mathematics and Information Networks (Beijing University of Posts and Telecommunications), Ministry of Education, China, under Grant No. KF202403.
\\
\\
$*$Corresponding author: sunwenrong@ustb.edu.cn

\nocite{}

\end{document}